\documentclass[a4paper,12pt,leqno]{article}
\usepackage{theorem}
\usepackage{latexsym,amssymb,amsfonts,amsmath}
\usepackage[dvips]{graphicx}
\usepackage{cite}
\newtheorem{thm}{Theorem}
\newtheorem{lem}{Lemma}

\theoremheaderfont{\scshape}

\title{{\Large {\bf Limit theorems for discrete-time quantum walks \\ on trees
}
}}
\author{ 
{\small Kota CHISAKI$^\sharp$,\quad 
Masatoshi HAMADA$^*$,\quad 
Norio KONNO$^\dagger$,\quad
Etsuo SEGAWA\footnote{To whom correspondence should be addressed. E-mail: segawa820@npde.osu.sci.ynu.ac.jp} }\\
{\scriptsize Department of Applied Mathematics, 
Faculty of Engineering, 
Yokohama National University
}\\
{\scriptsize Hodogaya, Yokohama 240-8501, Japan
} \\
{\scriptsize 
chisaki@npde.osu.sci.ynu.ac.jp$^\star$, mhamada@npde.osu.sci.ynu.ac.jp$^*$}, \\
{\scriptsize 
konno@ynu.ac.jp$^\dagger$, segawa820@npde.osu.sci.ynu.ac.jp$^\sharp$}\\
} 

\date{\empty }
\begin{document}
\maketitle

\par\noindent
\begin{small}
\par\noindent
{\bf Abstract}. We consider a discrete-time quantum walk $W_t$ given by the Grover transformation on the Cayley tree. 
We reduce $W_t$ to a quantum walk $X_t$ on a half line with a wall at the origin. 
This paper presents two types of limit theorems for $X_t$. 
The first one is $X_t$ as $t\to\infty$, which corresponds to a localization in the case of an initial qubit state. 
The second one is $X_t/t$ as $t\to\infty$, whose limit density is given by the Konno density function
~\cite{Konno1,Konno2,Miyazaki,Segawa}. 
The density appears in various situations of discrete-time cases. 
The corresponding 
similar limit theorem was proved in \cite{KonnoConti} for a continuous-time case on the Cayley tree.

\footnote[0]{
{\it Abbr. title:} Quantum walks on trees 
}

\end{small}

\setcounter{equation}{0}
\section{Introduction}
Let $G$ be a group generated by $\kappa\;(\geq 2)$ free involutions. 
The generating set is given by $\Sigma=\{\epsilon_1,\epsilon_2,\dots,\epsilon_{\kappa}\}$ 
with a relation $\epsilon_i^2=e$, where $e$ is the identity. 
The Cayley tree $\mathbb{T}_{\kappa}$ with the root $e$ is an infinite homogeneous $\kappa$-regular tree. 
The vertex set of $\mathbb{T}_\kappa$ is defined by the all possible reduced words in $G$ 
such that 
\[ V(\mathbb{T}_\kappa)
=\{\epsilon_{i_{n}}\epsilon_{i_{n-1}}\cdots\epsilon_{i_{1}}: 
\epsilon_{i_{j+1}}\neq \epsilon_{i_{j}}\;j=1,2,\dots,n\;\;(n\geq 1)\}\cup \{e\}. \]
Vertices $g$ and $h$ are connected if and only if $gh^{-1}\in \Sigma$. \\
\quad The state of a particle is described by a direct product of two 
Hilbert spaces $\mathcal{H}_P\otimes \mathcal{H}_C$, 
where $\mathcal{H}_P$ is generated by an orthonormal basis $\{|g\rangle;\;g\in V(\mathbb{T}_\kappa)\}$ and 
$\mathcal{H}_C$ is associated with an orthonormal basis $\{|\epsilon_j\rangle;\;\epsilon_j \in \Sigma\}$. 
The unitary time evolution $U$ is 
expressed as $U=S\cdot C$, where shift operator $S$ and coin operator $C$ 
act on a state $|\Psi\rangle\in \mathcal{H}_P\otimes \mathcal{H}_C$ 
in the following: 
if $|g,\epsilon\rangle$ is a base of $\mathcal{H}_P\otimes \mathcal{H}_C$, then 
\begin{align*}
C|g,\epsilon\rangle &= \sum_{\tau \in S}(-\delta_{\epsilon\tau}+2/\kappa)|g,\tau\rangle, \\
S|g,\epsilon\rangle &= |\epsilon g,\epsilon\rangle.
\end{align*}
Thus the one step unitary transition can be written as 
\begin{equation}\label{defQW}
U|g,\epsilon\rangle=\sum_{\tau \in \Sigma}(-\delta_{\epsilon\tau}+2/\kappa)|\tau g,\tau\rangle. 
\end{equation}
This paper is organized as follows. In Sect. 2, we reduce the quantum walk on $\mathbb{T}_\kappa$ to a walk on 
$\mathbb{Z}_+=\{0,1,2,\dots\}$. 
Section 3 presents two types of the limit theorems. 
Section 4 is denoted to summary and discussions. 
Appendixes A and B give proofs Theorems 1 and 2, respectively. 
\\
\\ 
\noindent
{\bf Acknowledgment.} 
We thank Takashi Oka for useful discussions and comments.
We also thank Takuya Machida for giving us nice figures (Figs. 1 and 2). 
\section{Reduction to half line}
Throughout this paper, we will consider the quantum walk starting from the root $e$ with 
the two cases of the initial qubit: for $\kappa\geq 3$, \\
\quad Case (A) Uniform initial qubit: $\varphi_0^{U}={}^T[1/\sqrt{\kappa},\dots, 1/\sqrt{\kappa}]$, \\
\quad Case (B) Weighted uniform initial qubit: 
$\varphi_0^{WU}={}^T[1/\sqrt{\kappa},\omega_{\kappa}/\sqrt{\kappa},\dots, \omega_\kappa^{\kappa-1}/\sqrt{\kappa}]$ 
with $\omega_{\kappa}=e^{2\pi i/\kappa}$. \\
Let us devide the set $V(\mathbb{T}_\kappa)\times \Sigma$ into a disjoint union of $A_j^{(\pm)}(x)$, 
$(j=0,1,\dots,\kappa-1,\;\; x\in\mathbb{Z}_+)$ with 
\begin{align*}
A^{(+)}_j(x) &= 
\begin{cases} \{ (e,\epsilon_j)\} & \text{:\;$x=0$,} \\
\{ (g,\epsilon)\in V(\mathbb{T}_{\kappa})\times \Sigma : |\epsilon g|=x+1, 
\mathrm{the}\; \mathrm{first}\; \mathrm{letter}\; \mathrm{of}\; g = \epsilon_j\} & \text{:\;$x\geq 1$,}
\end{cases} \\
A^{(-)}_j(x) &= 
\{ (g,\epsilon)\in V(\mathbb{T}_{\kappa})\times \Sigma : |\epsilon g|=x-1, 
\mathrm{the}\; \mathrm{first}\; \mathrm{letter}\; \mathrm{of}\; g = \epsilon_j \} \;\;\;\;:x\geq 1,
\end{align*}
where $|g|$ means the length of the reduced word $g$. 
To induce a reduction to a half line, 
we use the following lemma. 
\begin{lem}\label{reduction}
Let $\alpha_t(g,\epsilon)\in \mathbb{C}$ be probability amplitude at $(g,\epsilon)$ at time $t$, 
where $\mathbb{C}$ is the set of complex numbers. 
\begin{enumerate}
\item Case (A) (the initial qubit $\varphi_0^{U}$): \\ 
If $(g,\epsilon),(g',\epsilon')\in V(\mathbb{T}_\kappa)\otimes \Sigma$ with $|\epsilon g|=|\epsilon' g'|$, 
then $\alpha_t(g,\epsilon)=\alpha_t(g',\epsilon')$. 
\item Case (B) (the initial qubit $\varphi_0^{WU}$): \\ 
If $(g,\epsilon)\in A^{(\pm)}_i(x)$ and $(g',\epsilon')\in A^{(\pm)}_j(x)$, 
then $\alpha_t(g',\epsilon')=\omega_{\kappa}^{j-i}\alpha_t(g,\epsilon)$. 
\end{enumerate}
\end{lem}
\textit{Proof}. 
For part (1), 
from the symmetry of $\mathbb{T}_{\kappa}$ and the property of the Grover coin, 
we can show that for any $(g,\epsilon),(g',\epsilon')\in A^{(\pm)}_i(x)$, 
$\alpha_t(g,\epsilon)=\alpha_t(g',\epsilon')$ by induction on time step $t$, 
(see a more detailed proof in \cite{Kendon,Kendon2}, for example). 
Then when the initial qubit is $\varphi_0^U$, we see that if $(g,\epsilon),(g',\epsilon')\in V(\mathbb{T}_\kappa)\otimes \Sigma$ with 
$|\epsilon g|=|\epsilon' g'|$, then $\alpha_t(g,\epsilon)=\alpha_t(g',\epsilon')$.
For part (2), 
let $\mathcal{P}$ be a permutation on $\mathcal{H}_P\otimes \mathcal{H}_C$ such that 
for a basis $|\epsilon_{j_x}\cdots\epsilon_{j_1},\epsilon_m\rangle\in \mathcal{H}_P\otimes \mathcal{H}_C$, 
$\mathcal{P}|\epsilon_{j_x}\cdots\epsilon_{j_1},\epsilon_k\rangle
=|\epsilon_{j_x\oplus 1}\cdots\epsilon_{j_1\oplus 1},\epsilon_{k\oplus 1}\rangle$, where 
$x\oplus y=\mathrm{mod} \;[x+y,\kappa]$ and $\mathrm{mod}[a,b]$ is the remaider of $a/b$. 
We should note that if $(g,\epsilon)\in A_j^{(\tau)}(x)$ and 
$\mathcal{P}|g,\epsilon\rangle=|g',\epsilon'\rangle$, 
then $(g',\epsilon')\in A_{j\oplus 1}^{(\tau)}(x)$ ($\tau\in\{+,-\}$). 
The group generated by $\mathcal{P}$ is an automorphism group 
of $\kappa$-colored $\mathbb{T}_\kappa$ with color set $\Sigma$, i.e., $\mathcal{P}S\mathcal{P}^{-1}=S$, 
(see \cite{Krovi} for a detail). Then from the symmetry of the Grover coin, 
we have $\mathcal{P}U\mathcal{P}^{-1}=U$. 
Remark that the initial state 
$|e, \varphi_0^{WU}\rangle$ is the eigenvector of $\mathcal{P}$ with its eigenvalue $e^{-i\omega_\kappa}$. 
Let the total state at time $t$ be $|\Psi_t\rangle\equiv U^t|e,\varphi_0^{WU}\rangle$. 
Therefore we have $\mathcal{P}|\Psi_t\rangle = e^{-i\omega_\kappa}|\Psi_t\rangle$. \\

\noindent When the initial qubit is $\varphi_0^{U}$ or $\varphi_0^{WU}$, 
we can consider the time evolution under the subspace $\mathcal{H}'\subset \mathcal{H}_P\otimes \mathcal{H}_C$ 
generated by the following new basis: for the initial qubit $\varphi_0^{U}$, 
\begin{align*}
|x\rangle_{out} &= 
\frac{1}{\sqrt{\kappa (\kappa-1)^{x}}}\sum_{(g,\epsilon):|\epsilon g|=x+1}|g, \epsilon\rangle,
\;\;(x\geq 0), \\
|x\rangle_{in} &=  
\frac{1}{\sqrt{\kappa (\kappa-1)^{x-1}}}\sum_{(g,\epsilon): |\epsilon g|=x-1}|g, \epsilon\rangle,
\;\;(x\geq 1), 
\end{align*}
and, for the initial qubit $\varphi_0^{WU}$, 
\begin{align*}
|x\rangle_{out} &= 
\frac{1}{\sqrt{\kappa (\kappa-1)^{x}}}\sum_{j=0}^{\kappa-1}\omega_{\kappa}^{-j}\sum_{(\epsilon,g)\in A^{(+)}_j(x)}
|g, \epsilon\rangle,\;\;(x\geq 0), \\
|x\rangle_{in} &=  
\frac{1}{\sqrt{\kappa (\kappa-1)^{x-1}}}\sum_{j=0}^{\kappa-1}\omega_{\kappa}^{-j}\sum_{(\epsilon,g)\in A^{(-)}_j(x)}
|g, \epsilon\rangle,\;\;(x\geq 1).
\end{align*}
Therefore the one-step unitary transition defined by Eq. (\ref{defQW}) on the space $\mathcal{H}'$ is described as 
follows. If $\varphi_0\in\{\varphi_0^U, \varphi_0^{WU}\}$ be the initial qubit, then 
\begin{align}
U|x\rangle_{in} &= 
-(1-2/\kappa)|x-1\rangle_{out}+2\sqrt{\kappa-1}/\kappa |x+1\rangle_{in}\;\;\; :x\geq 1, \label{underH'1} \\ 
U|x\rangle_{out} &= 
\begin{cases}
|1\rangle_{in} & \text{:\;$x=0$, $\varphi_0=\varphi_0^{U}$,} \\
-|1\rangle_{in} & \text{:\;$x=0$, $\varphi_0=\varphi_0^{WU}$,} \\
(1-2/\kappa)|x+1\rangle_{in}+2\sqrt{\kappa-1}/\kappa |x-1\rangle_{out} 
		& \text{:\;$x \geq 1$, $\varphi_0\in\{\varphi_0^{U},\varphi_0^{WU}\}$. } \label{underH'2}
\end{cases}
\end{align}
Now we will show that the reduced quantum walk under a subspace $\mathcal{H}'$ with 
the time evolution given by Eqs.(\ref{underH'1}) and (\ref{underH'2}) 
is equivalent to 
a special case of quantum walk with a reflection wall at the origin on $\mathbb{Z}_+$ introduced by 
Oka \textit{et al}. \cite{Oka} in the following. 
At first we give the definition of the quantum walk with the wall. 
The space is described as 
$\widetilde{\mathcal{H}}_P\otimes\widetilde{\mathcal{H}}_C$, where $\widetilde{\mathcal{H}}_P$ is associated with 
an orthonormal basis 
$\{|x\rangle:x\in \mathbb{Z}\}$ and $\widetilde{\mathcal{H}}_C$ is generated by an orthonormal basis 
$\{|R\rangle,|L\rangle\}$. 
The time evolution $\widetilde{U}=\widetilde{S}\cdot \widetilde{C}$ 
on $\mathbb{Z}$ with the initial state $\Phi_0=|0,L\rangle$ is given by 
\begin{enumerate}
\item Coin operation: $\widetilde{C}|x,A\rangle=|x\rangle \otimes H(x)|A\rangle$ ($A=R,L$) with 
\begin{equation}\label{H(x)}
H(x)=(1-\delta_0(x))H_\kappa+e^{i\gamma}\delta_0(x)\sigma, 
\end{equation}
where $\gamma$ is a real number, $\delta_0(x)$ is the delta measure at the origin, 
\begin{equation*} 
H_{\kappa}=
\begin{bmatrix}2\sqrt{\kappa-1}/\kappa & -(1-2/\kappa) \\ 1-2/\kappa & 2\sqrt{\kappa-1}/\kappa \end{bmatrix}, 
\end{equation*}
and 
\[ \sigma=\begin{bmatrix}0&1\\1&0\end{bmatrix}. \]
\item Shift operation: 
\begin{equation*}\label{S}
\widetilde{S}|x,A\rangle=\begin{cases}|x+1,R\rangle & \text{: $(A=R)$}, \\ |x-1,L\rangle & \text{: $(A=L)$}. \end{cases}
\end{equation*}
\end{enumerate} 
Define a Hilbert space $\widetilde{\mathcal{H}'}$ 
as a subspace of $\widetilde{\mathcal{H}}_P\otimes\widetilde{\mathcal{H}}_C$ 
generated by a basis set 
\[ \{|0,L\rangle,|1,R\rangle,|1,L\rangle,|2,R\rangle,|2,L\rangle,\dots\}.\] 
Note that $\widetilde{U}^t|0,L\rangle \in \widetilde{\mathcal{H}'}$ for any $t\geq 0$. 
We should remark that the time evolution $U$ on $\mathcal{H}'$ given by Eqs.(\ref{underH'1}) and (\ref{underH'2}) 
is equivalent to the time evolution $\widetilde{U}$ on 
$\widetilde{\mathcal{H'}}$ with the following initial qubit $|L\rangle$ with the following one-to-one-correspondence: 
\begin{equation*}\label{1to1}
|x\rangle_{out} \leftrightarrow |x,L\rangle,\;\;|x\rangle_{in} \leftrightarrow |x,R\rangle. 
\end{equation*}
Furthermore 
the case of $\gamma=0$ (resp. $\gamma=\pi$) in Eq.(\ref{H(x)}) corresponds to the initial qubit 
$\varphi_0^{U}$ (resp. $\varphi_0^{WU}$). 
Let $W_t$ be the quantum walk on $\mathbb{T}_\kappa$ at time $t$ and $X_t$ be the quantum walk with the wall at time $t$. 
By definition, so we have $P(|W_t|=x)=P(X_t=x)$. 

\section{Limit theorems }
\begin{figure}[htb]
 \begin{minipage}{0.5\hsize}
  \begin{center}
   \includegraphics[width=30mm]{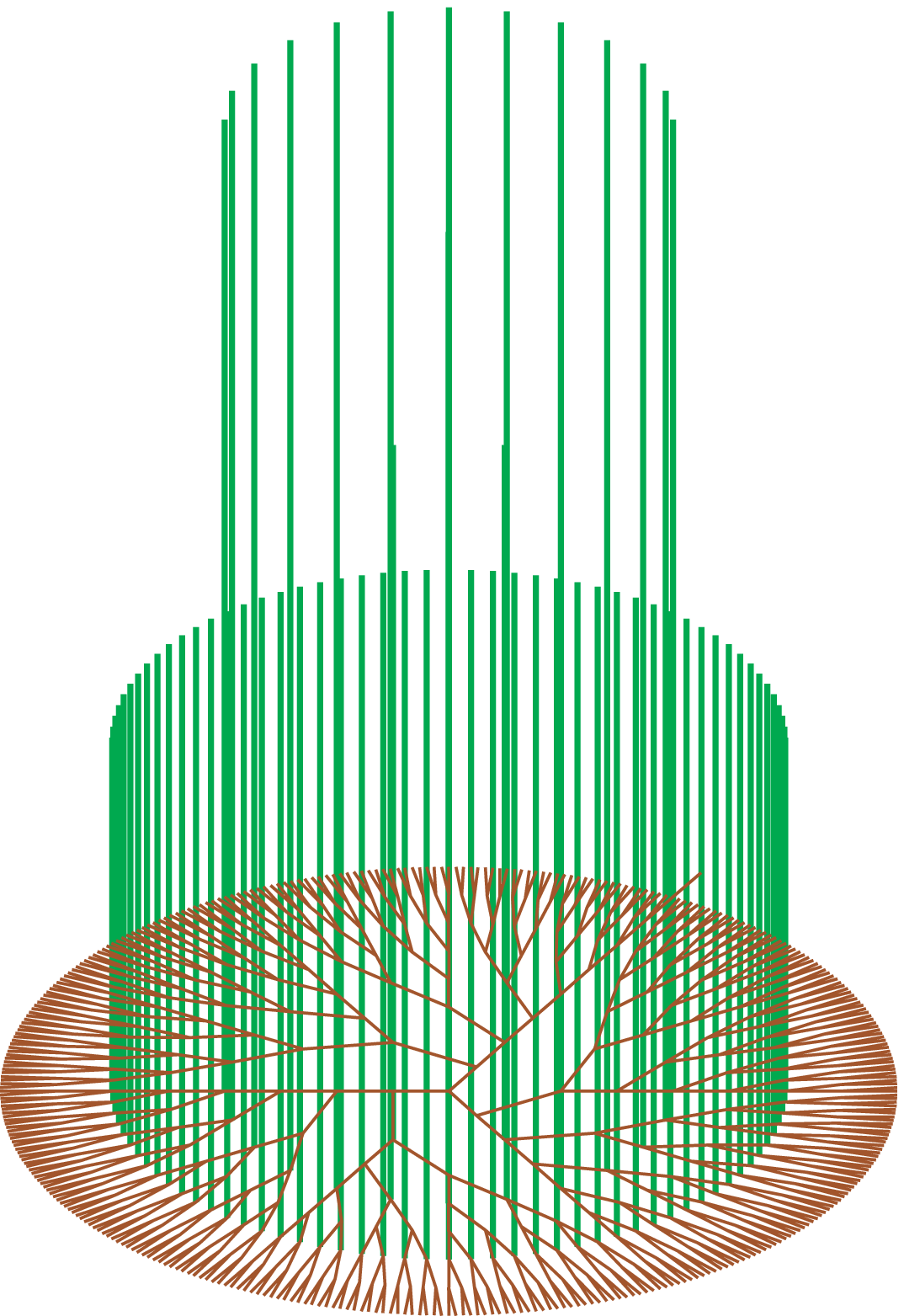}
  \end{center}
  \caption{Initial qubit is $\varphi_0^U$.}
  \label{fig:one}
 \end{minipage}
 \begin{minipage}{0.5\hsize}
  \begin{center}
   \includegraphics[width=30mm]{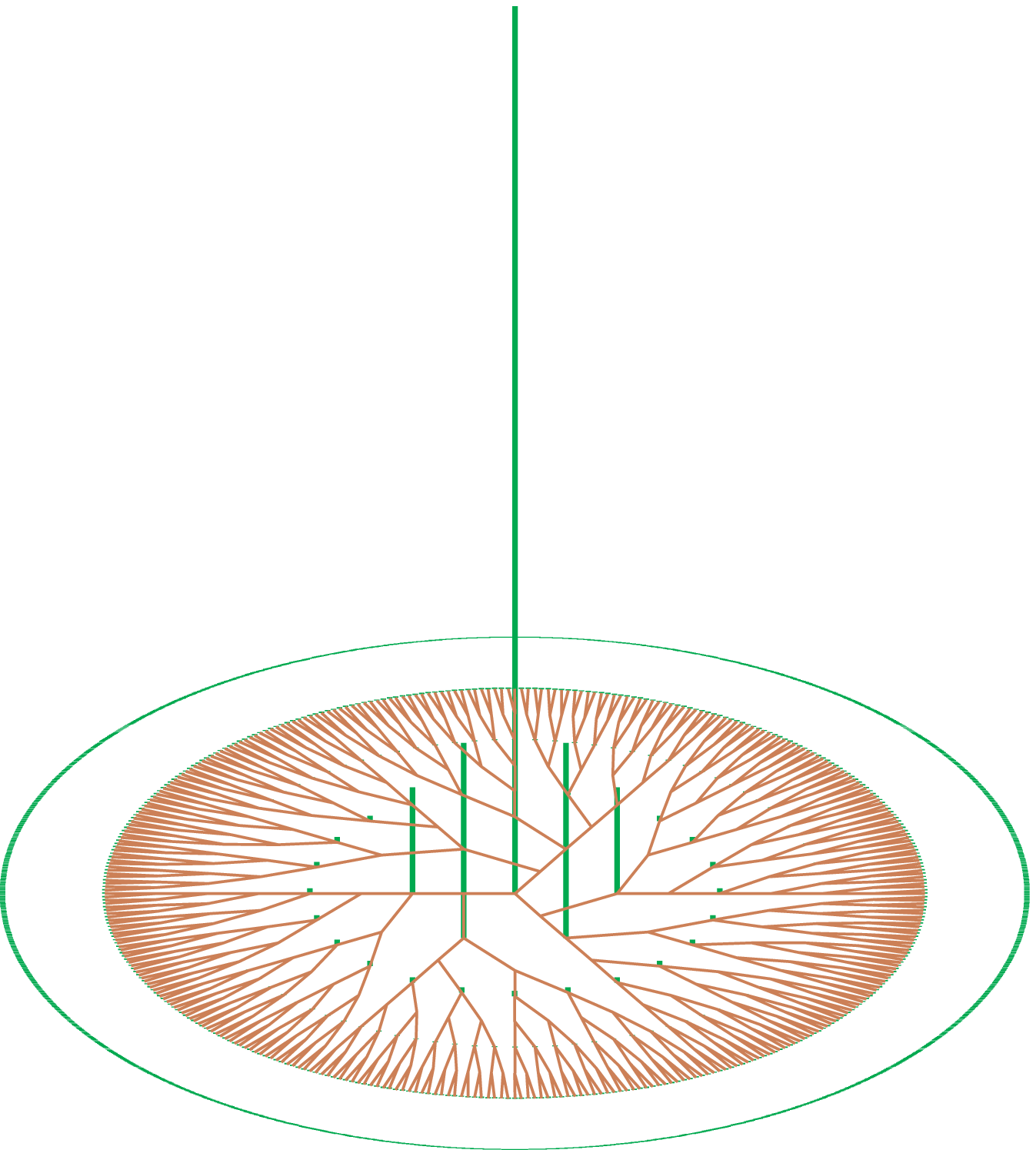}
  \end{center}
  \caption{Initial qubit is $\varphi_0^{WU}$.}
  \label{fig:two}
 \end{minipage}
\end{figure}

\begin{figure}[htb]
 \begin{minipage}{0.5\hsize}
  \begin{center}
   \includegraphics[width=40mm]{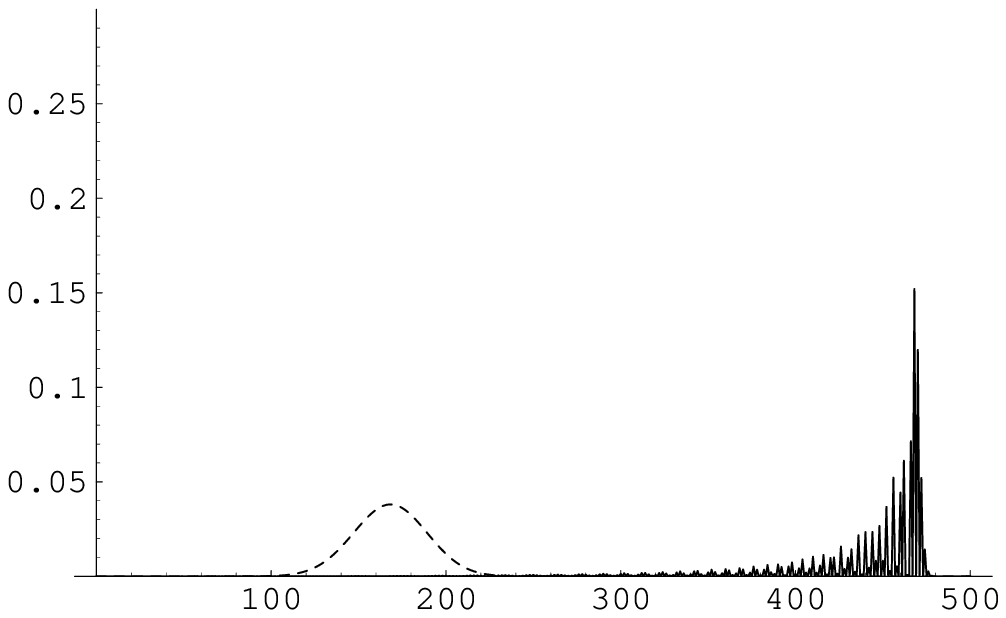}
  \end{center}
  \caption{Initial qubit is $\varphi_0^U$}
  \label{fig:three}
 \end{minipage}
 \begin{minipage}{0.5\hsize}
  \begin{center}
   \includegraphics[width=40mm]{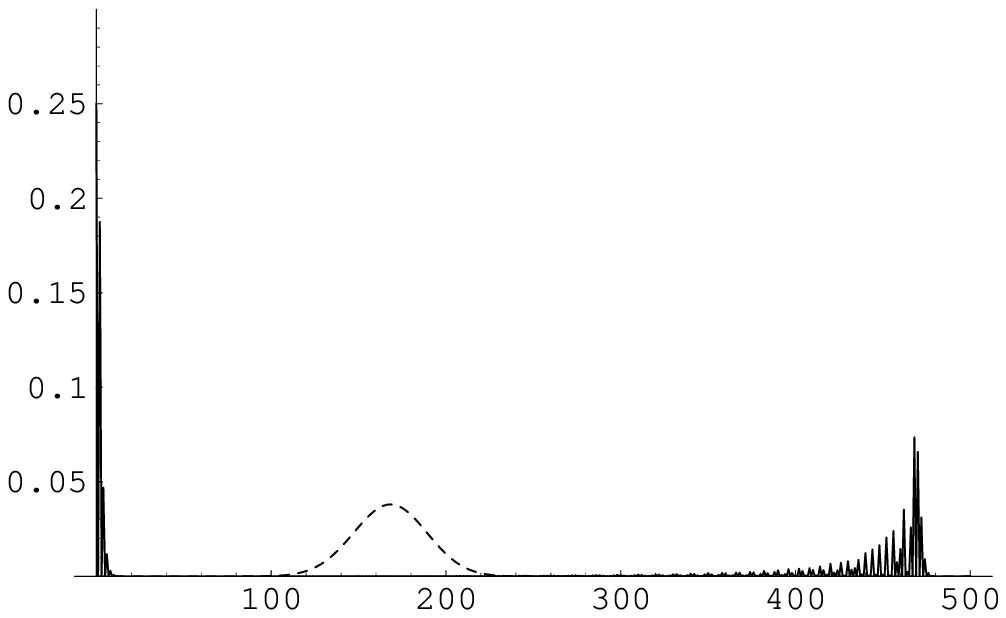}
  \end{center}
  \caption{Initial qubit is $\varphi_0^{WU}$}
  \label{fig:four}
 \end{minipage}
\end{figure}

In this section, we will show that a localization occurs in the case of the initial qubit $\varphi_0^{WU}$. 
The definition of the localization considered here is 
that there exists a vertex $v\in V(\mathbb{T}_\kappa)$ such that $\limsup_{t\to\infty}P(W_t=v)>0$. 
Figure \ref{fig:one} (resp. Fig. \ref{fig:two}) depicts 
the distribution of $W_t$ on $\mathbb{T}_3$ at time $t=10$ with the initial qubit $\varphi_0^U$ (resp. $\varphi_0^{WU}$). 
We can see that if $|g|=|h|$, then the finding probability at $g$ is equal to one at $h$ as 
we have shown in Lemma \ref{reduction}. 
Furthermore we can see a 
high probability at the origin with the initial qubit $\varphi_0^{WU}$. 
Figure. \ref{fig:three} (resp. Fig. \ref{fig:four}) shows the distribution of $X_t$ on $\mathbb{Z}_+$ 
at time 500 with the initial qubit $\varphi_0^U$ (resp. $\varphi_0^{WU}$). 
The solid lines in Figs. 3 and 4 
represent the quantum walk, and dotted lines in Figs. 3 and 4 represent the classical random walk. \\
\quad From now on, we present the limit theorems corresponding to a localization for $X_t$ and 
a weak convergence theorem for the rescaled $X_t/t$. 
The first theorem describes the localization for Case (B) suggested by Figs. 2 and 4. 
\begin{thm} \label{Localization}
Let $P_*^{(E)}(x)=\lim_{t\to\infty} P(X_{2t}=x)$ and $P_*^{(O)}(x)=\lim_{t\to\infty} P(X_{2t+1}=x)$ 
for $x\in \mathbb{Z}_+$. 
\begin{enumerate}
\item Case (A) ($\gamma=0$, i.e., $\varphi_0^{U}$ case) : for $x\in \mathbb{Z}_+$, 
\[ P_*^{(E)}(x)=P_*^{(O)}(x)=0\;\;(x\geq0 ). \]
\item Case (B) ($\gamma=\pi$, i.e., $\varphi_0^{WU}$ case)  : for $x\in \mathbb{Z}_+$, 
\begin{align*}
P_*^{(E)}(x)&= 
\begin{cases}
\left(\frac{\kappa-2}{\kappa-1}\right)^2
	\left\{\delta_0(x)+(1-\delta_0(x))\kappa\left(\frac{1}{\kappa-1}\right)^x\right\} & \text{; $x=even$,} \\
0 & \text{; $x=odd$.}        
\end{cases}\\
P_*^{(O)}(x)&= 
\begin{cases}
\kappa\left(\frac{\kappa-2}{\kappa-1}\right)^2
	\left(\frac{1}{\kappa-1}\right)^x & \text{; $x=odd$,} \\
0 & \text{; $x=even$.}        
\end{cases}
\end{align*}
\end{enumerate}
\end{thm}
The proof can be seen in Appendix A. 
Remark that for Case (A), $C_\kappa(A)\equiv \sum_{x\in\mathbb{Z}_+}P_*^{(E)}(x)=\sum_{x\in\mathbb{Z}_+}P_*^{(O)}(x)=0$, 
and for Case (B), 
\begin{equation*}\label{where} 
C_\kappa(B)\equiv \sum_{x\in\mathbb{Z}_+}P_*^{(E)}(x)=\sum_{x\in\mathbb{Z}_+}P_*^{(O)}(x)=\frac{\kappa-2}{\kappa-1}<1. 
\end{equation*}
That is, $\{P_*^{(E)}(x) : x\in\mathbb{Z}_+\}$ and $\{P_{*}^{(O)}(x): x\in\mathbb{Z}_+\}$ are 
not probability distributions for both cases. 
The following weak convergence theorem explains 
the vanishing values $1-C_\kappa(A)=1$ and $1-C_{\kappa}(B)=1/(\kappa-1)(>0)$. 
\begin{thm} \label{WCT}As $t\to\infty$, \\
\[ X_t/t \Rightarrow Y, \]
where ``$\Rightarrow$'' means the weak convergence. 
The limit measure is defined by 
\begin{equation} \label{Wthm}
\rho_\kappa (x) = 
\begin{cases}
f_{\kappa}(x) & \text{; Case (A),} \\
C_\kappa(B)\delta_0(x)+(1-C_\kappa(B))f_\kappa(x) & \text{; Case (B),}
\end{cases}
\end{equation}
where, 
\begin{equation} \label{KonnoW}
f_{\kappa}(x) = (\kappa-2) \frac{x^2 I_{[0,a_\kappa)}(x)}{\pi (1-x^2)\sqrt{a_\kappa^2-x^2}}, \;\;\;
a_\kappa=2\sqrt{\kappa-1}/\kappa,
\end{equation} 
and $I_A(x)$ is the indicator function of a set $A$. 
\end{thm}
As for the proof, see Appendix B. 
Note that the coefficient $\delta_0(x)$ in Eq. (\ref{Wthm}) 
for Case (B), i.e., $C_\kappa(B)$, corresponds to the localization. Furthermore 
$f_\kappa(x)$ in Eq.(\ref{KonnoW}) is described as the 
so-called Konno density function $\mu_K$ \cite{Konno1,Konno2} with a weight function $\kappa x^2$, that is, 
$f_\kappa(x)=\kappa x^2\mu_K(x,a_\kappa) I_{[0,\infty)}(x)$, 
where 
\[ \mu_K(x;a)=\frac{\sqrt{1-a^2}}{\pi(1-x^2)\sqrt{a^2-x^2}}I_{(-|a|,|a|)}(x). \]
The Konno density function 
appears in discrete-time quantum walks 
on $\mathbb{Z}$ \cite{Konno1, Konno2, Miyazaki} and 
on $\mathbb{Z}^2$ \cite{Watabe} as 
the limit density function for a suitable scaling, where $\mathbb{Z}$ is the set of integers. 
\section{Summary and discussions}
\quad 
We reduced a discrete-time quantum walk $W_t$ on the Cayley tree to a walk $X_t$ on $\mathbb{Z}_+$. 
We have obtained two types of limit theorems for $X_t$. 
The first one corresponds to a localization of $X_t$. 
The second one is a weak convergence theorem for $X_t/t$, 
where the limit density can be described by the Konno density function
~\cite{Konno1,Konno2,Miyazaki,Segawa}. 
To clarify a relation between the previous works of \cite{Jiang,Mirlin,Miller} and our result 
seems to be challenging. \\
\quad We can also reduce quantum walks on distance regular graphs such as the Hamming graph, 
the Johnson graph, etc., to a half line in a similar fashion. 
So the study on limit theorems for quantum walks on these graphs 
would be one of the future interesting problems. \\
\quad 
Finally we give an interesting relation 
between our discrete-time quantum walk on $\mathbb{T}_\kappa$ 
and the continuous-time quantum walk on $\mathbb{T}_\kappa$ studied by \cite{KonnoConti} 
with respect to the weak convergence. 
The total Hilbert space of the continuous-time quantum walk on $\mathbb{T}_\kappa$ 
is associated with an orthonormal basis $\{|g\rangle: g\in V(\mathbb{T}_\kappa)\}$. The 
state $\Psi_t^{(c)}$ at time $t$ with the initial state $|e\rangle$ is given by $\Psi^{(c)}_t\equiv U^t|e\rangle$ with 
$U_t=e^{itA_\kappa/\sqrt{\kappa}}$, where $A_\kappa$ is the adjacency matrix of $\mathbb{T}_\kappa$, i.e., 
$(A_\kappa)_{g,h}=I_{\{(g,h): gh^{-1}\in \Sigma\}}(g,h)$. Here $I_X(x,y)=1$, if $(x,y)\in X$, $=0$, if $(x,y)\notin X$. 
Let $|x\rangle=|e\rangle$ $(x=0)$, $=1/\sqrt{\kappa(\kappa-1)^{x-1}}\sum_{g:|g|=x}|g\rangle$ ($x\geq 1$).  
Then we can reduce 
the continuous-time quantum walk on $\mathbb{T}_\kappa$ to a walk on the subspace ${\mathcal{H}^{(c)}}'$ generated by 
$\{|x\rangle : x\in \mathbb{Z}_+\}$ as in the discrete-time case. 
Assume that $\alpha_t(x)$ denotes the amplitude at time $t$ 
at position $x$ of the reduced walk on $\mathbb{Z}_+$. 
By a quantum probabilistic approach \cite{Salimi,Obata}, the following limit theorem was shown in \cite{KonnoConti}: 
\[\lim_{\kappa\to\infty}\alpha_t(x)=(x+1)i^x\frac{J_{x+1}(2t)}{t}, \]
where $J_{x}(n)$ denotes the Bessel function of the first kind of order $n$. 
Let $X^{(c)}_t$ be a continuous-time quantum walk starting from the origin defined by 
\[ P(X_t^{(c)}=x)= (x+1)^2\frac{J_{x+1}^2(2t)}{t^2}. \]
Furthermore, the following weak limit theorem was proved in \cite{KonnoConti}:
as $t\to \infty$, 
\[ X_t^{(c)}/t \Rightarrow Y^{(c)}, \]
where $Y^{(c)}$ has the density 
\[ \rho^{(c)}(x)=x^2\mu_A(x;2)I_{[0,\infty)}(x),  \]
with $\mu_A(x;a)=I_{(-|a|,|a|)}(x)/(\pi\sqrt{a^2-x^2})$. 
As shown in \cite{KonnoContiZ}, 
$\mu_A(x;a)$ is the rescaled limit density function for a \textit{continuous-time} quantum walk on $\mathbb{Z}$. 
On the other hand, Theorem 2 gives a similar result: 
\[ f_\kappa(x)=\kappa x^2 \mu_K(x;a_\kappa)I_{[0,\infty)}(x),  \]
where the Konno density function $\mu_K$ is the rescaled limit density function for 
a \textit{discretei-time} quantum walk on $\mathbb{Z}$. 
\quad \\
\quad \\
\noindent
{\bf Appendix A: Proof of Theorem \ref{Localization}} \\

Let $\Psi_t(x)$ be the coin state at time $t$ and position $x\in\mathbb{Z}_+$ 
of the quantum walk with the reflection wall at the origin. 
Let $\widetilde{\Psi}(x;z)$ denote a generating function for $\Psi_t(x)$ such that 
$\widetilde{\Psi}(x;z)=\sum_{t\geq 0}\Psi_{t}(x)z^t$. From the result of \cite{Oka}, 
we can obtain an explicit expression for 
$\widetilde{\Psi}(x;z)={}^T[\widetilde{\Psi}^{(L)}(x;z),\widetilde{\Psi}^{(R)}(x;z)]$ in the following: 
\begin{align}
\widetilde{\Psi}^{(L)}(x;z) &= \begin{cases} m_{\kappa}(\lambda(z))^{x-1}(a_{\kappa}z-\lambda(z))\frac{\nu(z)}{z^2-1} & \text{;\;$x \geq 1$,} \\
 				       m_{\kappa}(z-a_{\kappa}\lambda(z))z\frac{\nu(z)}{z^2-1} & \text{;\;$x=0$,}	\end{cases} \label{oka}\\
\widetilde{\Psi}^{(R)}(x;z) &= \begin{cases} z(\lambda(z))^{x-1}\frac{\nu(z)}{z^2-1} & \text{;\;$x \geq 1$,} \\
					     0	& \text{;\;$x=0$,}\end{cases} \label{oka2}
\end{align}
with $a_{\kappa}=2\sqrt{\kappa-1}/\kappa$,  $m_{\kappa}=\kappa/(\kappa-2)$ (Case A), $=-\kappa/(\kappa-2)$ (Case B), and 
\begin{align*}
\lambda(z) &= \frac{z^2+1-\sqrt{z^4+2(1-2a_\kappa^2)z^2+1}}{2a_\kappa z}, \\
\nu(z) &= \frac{2-m_{\kappa}+m_{\kappa} z^2-m_{\kappa}\sqrt{z^4+2(1-2a_\kappa^2)z^2+1}}{2(1-m_{\kappa})}.
\end{align*}
For $r_0\in (0,1)$, we get 
\[ \Psi_t(x)=\frac{1}{2\pi i}\int_{|z|=r_0} \widetilde{\Psi}(x;z)\frac{dz}{z^{t+1}}. \]
Remark that $||\widetilde{\Psi}(x;z)||^2<1$. Then $\int_{|z|=r_1}\widetilde{\Psi}(x;z)/z^{t+1} dz \to 0$ 
with $r_1>1$ as $t\to\infty$. So we have 
\begin{equation*}\label{reason} 
\Psi_t(x) \to -\left(\mathrm{Res}(\widetilde{\Psi}(x;z),1)+\mathrm{Res}(\widetilde{\Psi}(x;z),-1)(-1)^{t+1}\right) \;\;(t \to \infty). 
\end{equation*}
The above equation gives 
\begin{align*}
\lim_{t\to\infty}\Psi^{(L)}_t(x) 
	&= \begin{cases} (1+(-1)^{x+t})\;m_\kappa\;\nu(1) 
        \frac{\kappa-2}{2\kappa \sqrt{\kappa-1}}\left(\frac{1}{\sqrt{\kappa-1}}\right)^{x-1} & \text{;\;$x \geq 1$,} \\
 	(1+(-1)^{x+t})\;m_\kappa\;\nu(1)(1/2-1/\kappa) & \text{;\;$x=0$,}	\end{cases} \\
\lim_{t\to\infty}\Psi^{(R)}_t(x) 
	&= \begin{cases} (1+(-1)^{x+t})\;\nu(1)
        \left(\frac{1}{\sqrt{\kappa-1}}\right)^{x-1}/2 & \text{;\;$x \geq 1$,} \\
					     0	& \text{;\;$x=0$.} \end{cases}        
\end{align*} 
The desired conclusion follows from 
$\nu(1)=0$ (Case (A)), $=(\kappa-2)/(\kappa-1)$ (Case (B)). \\

\par\noindent
{\bf Appendix B: Proof of Theorem \ref{WCT}} \\

Let $\widehat{\widetilde{\Psi}}(k;z)={}^T[\widehat{\widetilde{\Psi}}^{(L)}(k;z),\widehat{\widetilde{\Psi}}^{(R)}(k;z)]
=\sum_{x}\widetilde{\Psi}(x;z)e^{ikx}$. When $|\lambda(z)|<1$, Eqs. (\ref{oka}) and (\ref{oka2}) imply
\begin{align*} 
\widehat{\widetilde{\Psi}}^{(L)}(k;z)
	&=\frac{\phi_0^{(L)}(k,z)}{(z+1)(z-1)}
        +\frac{\phi_1^{(L)}(k,z)}{z(z+1)(z-1)(z-e^{i\theta(k)})(z-e^{-i\theta(k)})}, \\
\widehat{\widetilde{\Psi}}^{(R)}(k;z)
	&=\frac{\phi^{(R)}(k,z)}{(z+1)(z-1)(z-e^{i\theta(k)})(z-e^{-i\theta(k)})},
\end{align*}
where $\phi_0^{(L)}(k,z)$, $\phi_1^{(L)}(k,z)$, $\phi^{(R)}(k,z)$ are some regular functions on $\mathbb{C}$, and 
$\cos\theta(k)=a_\kappa\cos k$. 
Since $||\widehat{\widetilde{\Psi}}(k;z)||^2<\infty$ ($|z|<1$), we can rewrite 
$\widehat{\widetilde{\Psi}}(k;z)$ as $\widehat{\widetilde{\Psi}}(k;z)=\sum_{t\geq 0}\widehat{\Psi}_t(k)z^t$ with 
$\widehat{\Psi}_t(k)=\sum_{x}\Psi_t(x)e^{ikx}$. 
For $r_0\in (0,1)$, we have
\[ \widehat{\Psi}_t(k)=\frac{1}{2\pi i}\int_{|z|=r_0}\widehat{\widetilde{\Psi}}(k,z)\frac{dz}{z^{t+1}}.  \]
Then for $|z|>1$, $||\widehat{\widetilde{\Psi}}(k,z)||^2<\infty$ implies 
$\int_{|z|=r_1}\widehat{\widetilde{\Psi}}(k,z)\frac{dz}{z^{t+1}}\to 0$  ($t \to \infty$) with $r_1>1$. 
So 
\begin{equation*} 
-\widehat{\Psi}_t(k) \to 
	        \psi_1(k)+\psi_{-1}(k)(-1)^{t+1}+\psi_{+}(k)e^{-i(t+1)\theta(k)}+\psi_{-}(k)e^{i(t+1)\theta(k)}
                \;\;(t\to\infty),
\end{equation*}
where $\psi_{\pm 1}(k)=\mathrm{Res}(\widehat{\widetilde{\Psi}}(k,z);\pm 1)$ and 
$\psi_{\pm}(k)=\mathrm{Res}(\widehat{\widetilde{\Psi}}(k,z);e^{\pm i\theta(k)})$. 
The definition of $\widehat{\Psi}_t(k)$ gives 
\begin{equation}\label{chara}
E\left[e^{i\xi X_t}\right] = \int_{0}^{2\pi} \langle \widehat{\Psi}_t(k),\widehat{\Psi}_t(k+\xi) \rangle \frac{dk}{2\pi}.
\end{equation}
Hence  
\begin{align}
\int_{0}^{2\pi}\left(||\psi_1(k)||^2+||\psi_{-1}(k)||^2\right)\frac{dk}{2\pi} &= (\kappa-2)/(\kappa-1), \label{RHS}\\
\int_{0}^{2\pi}\left(\langle \psi_1(k), \psi_{-1}(k)\rangle+\langle \psi_{-1}(k), \psi_{1}(k)\rangle\right) 
	\frac{dk}{2\pi} &=0. \label{zero}
\end{align}
Note that the right-hand side of Eq. (\ref{RHS}) is nothing but $C_\kappa(B)$. 
Combining Eqs. (\ref{chara}), (\ref{RHS}) and (\ref{zero}) with the Riemann-Lebesgue lemma, we have 
\begin{equation}\label{main}
\lim_{t \to \infty} E\left[e^{i\xi X_t/t}\right]
	 = \frac{\kappa-2}{\kappa-1}+\int_0^{2\pi} e^{-i\xi h(k)} p(k)\frac{dk}{2\pi}+\int_0^{2\pi} e^{i\xi h(k)} q(k)\frac{dk}{2\pi},
\end{equation}
where $h(k)=d\theta(k)/dk$, $p(k)=||\psi_{+}(k)||^2$, $q(k)=||\psi_{-}(k)||^2$. 
An explicit expression for $h(k)$ is 
\[ h(k)=a_\kappa\sin{k}/\sqrt{1-a_\kappa^2\cos^2 k}.\] 
Then $h'(k)\equiv dh(k)/dk=a_\kappa(1-a_\kappa^2)\cos k/(1-a_\kappa^2\cos^2 k)^{3/2}$. 
Put $h_{+}(k)=I_{[0,\pi/2)\cup [3\pi/2,2\pi)}(k)h(k)$ and $h_{-}(k)=I_{(\pi/2,3\pi/2)}(k)h(k)$. 
If $x=h_{\pm}(k)$ with $|x|\leq a_\kappa$, then the solutions $k_\pm(x)$ are given by 
\begin{equation*} 
	\cos (k_\pm(x)) = \pm\frac{1}{a_\kappa}\sqrt{\frac{a_\kappa^2-x^2}{1-x^2}}, \;\;
        \sin (k_\pm(x)) = \frac{\sqrt{1-a_\kappa^2}}{a_\kappa} \frac{x}{\sqrt{1-x^2}}.
\end{equation*} 
Therefore we obtain 
\begin{align*}
h'(k_{\pm}(x)) &= \pm \frac{(1-x^2)\sqrt{a_\kappa^2-x^2}}{\sqrt{1-a_\kappa^2}}, \\
p(k_{\pm}(x))  &= (1+\mathrm{sgn}(x))\frac{(\kappa-2)^2}{4\kappa(\kappa-1)}\frac{x^2}{a_\kappa^2-x^2}, \\         
q(k_{\pm}(x))  &= (1+\mathrm{sgn}(x))\frac{(\kappa-2)^2}{4\kappa(\kappa-1)}\frac{x^2}{a_\kappa^2-x^2},  
\end{align*}
where $\mathrm{sgn}(x)=1$ $(x>0)$, $=0$ $(x=0)$, $=-1$ $(x<0)$. 
Then by putting $h(k)=x$, the second and third terms of right-hand side of Eq. (\ref{main}) can be expressed as 
\[\int_{0}^{2\pi}\left(e^{-i\xi h(k)} p(k)+e^{i\xi h(k)} q(k)\right)\frac{dk}{2\pi}
	=\int_{0}^{\infty}e^{i\xi x}w(x)\mu_K(x;a_\kappa)dx, \]
where $\mu_K(x;a)$ is the Konno density function and weight function $w(x)$ is given by  
\begin{equation*}\label{weight}
w(x)=\begin{cases}\kappa x^2 & \text{; Case (A),} \\ 
\frac{\kappa}{\kappa-1}x^2 & \text{; Case (B).}
\end{cases}  
\end{equation*}
Thus we obtain the desired conclusion.

\begin{small}
\bibliographystyle{jplain}

\begin{thebibliography}{99}




\bibitem{Konno1}
Konno, N., 
``Quantum random walks in one dimension,'' 
\textit{Quantum Information Processing}, \textbf{1}: 345-354 (2002). 

\bibitem{Konno2}
Konno, N., 
``A new type of limit theorems for the one-dimensional quantum random walk,'' 
\textit{Journal of the Mathematical Society of Japan}, \textbf{57}: 1179-1195 (2005). 

\bibitem{Miyazaki}
Miyazaki, T., Katori, M., and Konno, N., 
``Wigner formula of rotation matrices and quantum walks,'' 
\textit{Physical Review A}, \textbf{76}: 012332 (2007). 

\bibitem{Segawa}
Segawa, E., and Konno, N., 
``Limit theorems for quantum walks driven by many coins,'' 
\textit{International Journal of Quantum Information} \textbf{6}: 1231-1243 (2008). 

\bibitem{KonnoConti}
Konno, N., 
``Continuous-time quantum walks on trees in quantum probability theory,'' 
\textit{Infinite Dimensional Analysis, Quantum Probability and Related Topics}, \textbf{9}: 287-297 (2006). 

\bibitem{Kendon}
Tregenna, B., Flanagan, W., Maile, R., and Kendon, V., 
``Controlling discrete quantum walks: coins and initial states,'' 
\textit{New Journal of Physics}, \textbf{5}: 83 (2003). 

\bibitem{Kendon2}
Carneiro, I., Loo, M., Xu, X., Girerd, M., Kendon, V., and Knight, P. L., 
``Entanglement in coined quantum walks on regular graphs,'' 
\textit{New Journal of Physics}, \textbf{7}: 156 (2005) . 

\bibitem{Krovi}
Krovi, H., and Brun, T. A., 
``Quantum walks on quotient graphs,'' 
\textit{Physical Review A}, \textbf{75}: 062332 (2007). 


\bibitem{Oka}
Oka, T., Konno, N., Arita, R., and Aoki, H., 
``Breakdown of an electric-field driven system: a mapping to a quantum walk,'' 
\textit{Physical Review Letter}, \textbf{94}: 100602 (2005).

\bibitem{Watabe}
Watabe, K., Kobayashi, N., Katori, M., and Konno, N., 
``Limit distributions of two-dimensional quantum walks,'' 
\textit{Physical Review A}, \textbf{77}: 062331 (2008). 

\bibitem{Jiang}
Jiang, D. L., and Aida, T., 
``Photoisomerization in dendrimers by harvesting of low-energy photons,'' 
\textit{Nature}, \textbf{388}: 454-456 (1997). 

\bibitem{Mirlin}
Mirlin, A. D., and Fyodorov, Y. V., 
``Localization transition in the anderson model on the bethe lattice: spontaneous symmetry 
breaking and correlation functions,'' 
\textit{Nuclear Physics B}, \textbf{366}: 507-532 (1991). 

\bibitem{Miller}
Miller, J. D., and Derrida, B., 
``Weak-disorder expansion for the Anderson model on a tree,'' 
\textit{Journal of Statistical Physics}, \textbf{75}: 357-388 (1994).

\bibitem{Salimi}
Jafarizadeh, M. A., and Salimi, S., 
``Investigation of continuous-time quantum walk via spectral distribution associated with adjacency matrix,'' 
\textit{Annals of Physics}, \textbf{322}: 1005-1033 (2007). 

\bibitem{Obata}
Obata, N., 
``Quantum probabilistic approach to spectral analysis of star graphs,'' 
\textit{Interdisciplinary Information Sciences}, \textbf{10}: 41-52 (2004).

\bibitem{KonnoContiZ} 
Konno, N., 
``Limit theorem for continuous-time quantum walk on the line,'' 
\textit{Physical Review E}, \textbf{72}: 026113 (2005).















\end{thebibliography}

\end{small}

\end{document}